\documentclass[11pt,a4paper]{emulateapj}
\usepackage[dvipdfmx]{hyperref}

\usepackage{epsfig}
\usepackage{amsmath}
\usepackage{natbib}
\usepackage{multirow}

\def\arcsec{\hbox{$^{\prime\prime}$}}
\def\deg{\hbox{$^\circ$}}

\begin{document}

\title{On the Merging Cluster Abell\,578 and Its Central Radio Galaxy 4C\,+67.13\footnotemark[1]}
\footnotetext[1]{Based on service observations made with the WHT operated on the island of La Palma by the Isaac Newton Group in the Spanish Observatorio del Roque de los Muchachos of the Instituto de Astrof\'isica de Canarias.}

\author{K.~Hagino\altaffilmark{1,2}, \L.~Stawarz\altaffilmark{1,3}, A.~Siemiginowska\altaffilmark{4}, C.C.~Cheung\altaffilmark{5}, D.~Kozie\l-Wierzbowska\altaffilmark{3}, A.~Szostek\altaffilmark{3}, G.~Madejski\altaffilmark{6}, D.E.~Harris\altaffilmark{4},  A.~Simionescu\altaffilmark{1}, and T.~Takahashi\altaffilmark{1,2}}
\affil{
\altaffilmark{1}Institute of Space and Astronautical Science JAXA, 3-1-1 Yoshinodai, Chuo-ku, Sagamihara, Kanagawa 252-5210, Japan\\
\altaffilmark{2}Department of Physics, University of Tokyo, 7-3-1 Hongo, Bunkyo, Tokyo 113-0033, Japan\\
\altaffilmark{3}Astronomical Observatory, Jagiellonian University, ul. Orla 171, 30-244 Krak\'ow, Poland\\
\altaffilmark{4}Harvard Smithsonian Center for Astrophysics, 60 Garden St, Cambridge, MA 02138, USA\\
\altaffilmark{5}Space Science Division, Naval Research Laboratory, Washington, DC 20375, USA\\
\altaffilmark{6}W. W. Hansen Experimental Physics Laboratory, Kavli Institute for Particle Astrophysics and Cosmology, Department of Physics\\
and SLAC National Accelerator Laboratory, Stanford University, Stanford, CA 94305, USA
}

\email{email: {\tt hagino@astro.isas.jaxa.jp}}

\begin{abstract}

Here we analyze radio, optical, and X-ray data for a peculiar cluster Abell\,578. This cluster is not fully relaxed and consists of two merging sub-systems. The brightest cluster galaxy, CGPG\,0719.8+6704, is a pair of interacting ellipticals with projected separation $\sim 10$\,kpc, the brighter of which hosts the radio source 4C\,+67.13. The Fanaroff-Riley type-II radio morphology of 4C\,+67.13 is unusual for central radio galaxies in local Abell clusters. Our new optical spectroscopy revealed that both nuclei of the CGPG\,0719.8+6704 pair are active, 
albeit at low accretion rates corresponding to the Eddington ratio $\sim10^{-4}$ (for the estimated black hole masses of $\sim 3 \times 10^8 \, M_{\odot}$ and $\sim 10^9 \, M_{\odot}$). The gathered X-ray ({\it Chandra}) data allowed us to confirm and to quantify robustly the previously noted elongation of the gaseous atmosphere in the dominant sub-cluster, as well as a large spatial offset ($\sim 60$\,kpc projected) between the position of the brightest cluster galaxy and the cluster center inferred from the modeling of the X-ray surface brightness distribution. Detailed analysis of the brightness profiles and temperature revealed also that the cluster gas in the vicinity of 4C\,+67.13 is compressed (by a factor of about $\sim 1.4$) and heated (from $\simeq 2.0$\,keV up to 2.7\,keV), consistent with the presence of a weak shock (Mach number $\sim 1.3$) driven by the expanding jet cocoon. This would then require the jet kinetic power of the order of $\sim 10^{45}$\,erg\,s$^{-1}$, implying either a very high efficiency of the jet production for the current accretion rate, or a highly modulated jet/accretion activity in the system.
\end{abstract}


\section{Introduction}

Relativistic jets produced in active galactic nuclei (AGN) are thought to play a major role in shaping the co-evolution of supermassive black holes (SMBHs) and galaxies via widely discussed, though still hardly understood, feedback processes \citep[see the topical reviews by][]{fab12,kor13}. The most direct manifestation of the impact the jets make on their environment is the presence of cavities inflated in the intergalactic/intracluster medium at the position of jet cocoons (or `radio lobes'), which are often accompanied by shock waves driven in the surrounding hot, X-ray emitting gas \citep[see, e.g.,][and references therein]{mac12}.

The more luminous the jets are, the more dramatic their influence is on the ambient medium. It has been argued in particular, that high-power jets are necessary to explain the observed properties of the central parts of rich clusters as due to mechanical gas heating by expanding lobes \citep[see, e.g.,][]{voi05,mat11}. These high-powers are equivalent to those in radio-loud quasars and their parent population of Fanaroff-Riley type-II radio galaxies (hereafter FR\,IIs). Yet, most of the jetted AGN found in the centers of nearby clusters are low-power systems, with large-scale edge-dimmed FR\,I radio morphologies \citep[e.g.,][]{zir97,win11}. One notable exception from this rule is the FR\,II radio galaxy Cygnus~A, located in the center of a rich cluster \citep{car96}. This source is considered to be representative of high-redshift radio galaxies since in the earlier Universe, when the bulk of the feedback process took place, powerful AGN with luminous jets are being found also in dense environments \citep[e.g.,][]{sie05,bel07,ant12}

\begin{figure*}[tbp]
\begin{center}
\includegraphics[width=0.9\columnwidth]{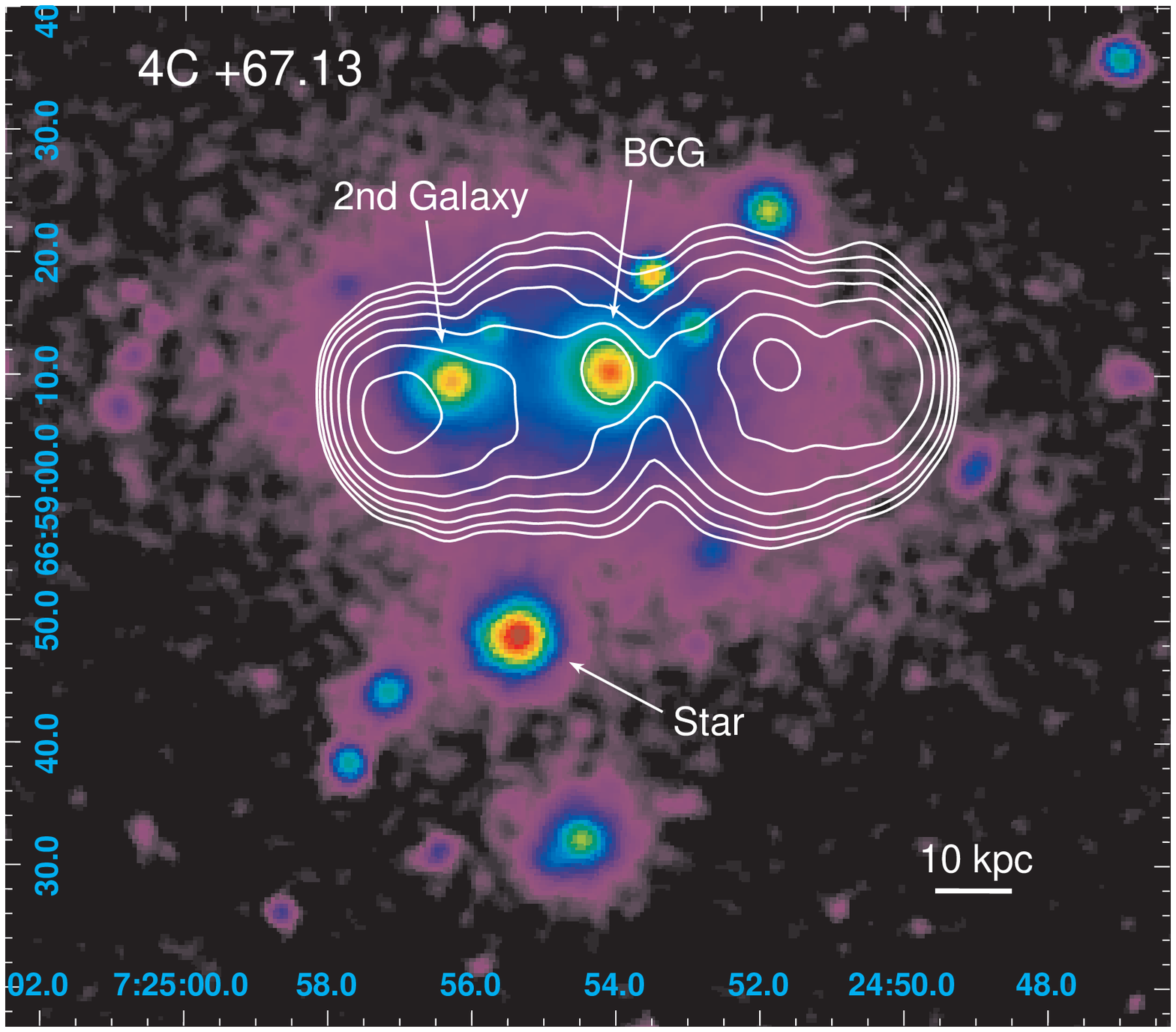}
\includegraphics[width=0.9\columnwidth]{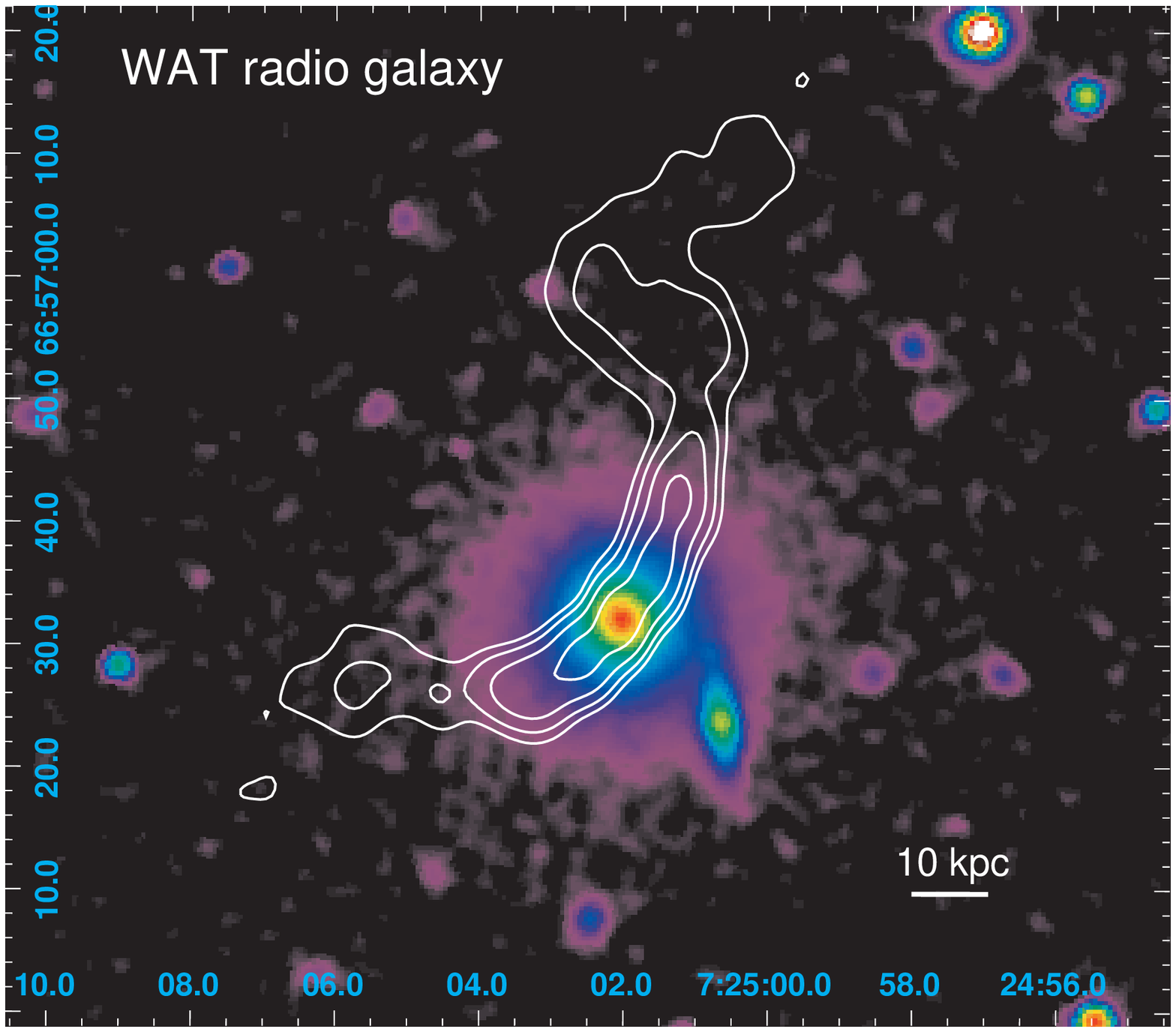}
\caption{VLA 4.7\,GHz contours (levels start at 0.2 mJy beam$^{-1}$ and increase by factors of two) of the two radio galaxies found in the Abell\,578 cluster superimposed on smoothed optical SDSS $i$-band images. The left panel shows the BCG-hosted 4C\,+67.13, and the right panel shows the WAT radio galaxy located in the southern part of the cluster. The fainter second galaxy adjacent to the 4C\,+67.13 host, which is seen through the eastern radio lobe (in projection), is the other member of the galaxy pair CGPG\,0719.8+6704. A bright optical source visible to the south of 4C\,+67.13 is a foreground star. Throughout, coordinates are in J2000.0 equinox.}
\label{VLAimage}
\end{center}
\end{figure*}

In order to increase the sample of FR\,II radio sources studied in nearby clusters, we examined the NRAO\footnote{The National Radio Astronomy Observatory is a facility of the National Science Foundation operated under cooperative agreement by Associated Universities, Inc.} Very Large Array (VLA) imaging survey of Abell clusters with richness $R \geq 0$ \citep{owe92,owe97}. We found that within the redshift range $z < 0.25$ and at the northern declinations visible to the VLA, only two FR\,II radio galaxies are hosted by brightest cluster galaxies (BCGs) in this sample: PKS\,B1358$-$113 in Abell\,1836 and 4C\,+67.13 in Abell\,578. For these two systems, we have re-analyzed archival VLA maps, acquired new optical spectra using the William Herschel Telescope (WHT), and obtained new X-ray data with Chandra and XMM-Newton. In this paper we present the analysis\footnote{The XMM-Newton data we obtained in this case was entirely contaminated by background flares and is not considered in this paper.} of 4C\,+67.13/Abell\,578, which is a peculiar example of a cluster in formation. The companion study of PKS\,B1358$-$113/Abell\,1836 is presented in \citet{sta14}.

Based on a detailed analysis of the velocity measurements for the cluster member galaxies, along with ROSAT PSPC data, \citet{gom97} concluded that Abell\,578 (cluster richness $R = 0$) is not relaxed, and consists of two interacting sub-clusters with a velocity difference of $\sim1300$~km~s$^{-1}$. The BCG is associated with the dominant cluster, and is the brighter of a pair of interconnected red spherical compacts as noted by \citet{zwi71} and catalogued by him as CGPG 0719.8+6704. The radio galaxy 4C\,+67.13 is hosted by the BCG and is characterized by an FR\,II large-scale morphology \citep[][designated as 0719+670 therein]{owe97}, and hence potentially by a high jet kinetic power. At the southern end of this dominant cluster is the sub-cluster containing another radio galaxy, designated as 0720+670 by \citet{owe97}, which is of the `wide-angle tailed' (WAT) type (see Section~\ref{S-radio}). Both radio galaxies are at rest with respect to their sub-clusters \citep{gom97}.

Below, we assume a standard cosmology with $H_0=71$\,km\,s$^{-1}$\,Mpc$^{-1}$, $\Omega_{\rm m}=0.27$ and $\Omega_\Lambda=0.73$, so that the redshift of the target $z = 0.0866$ \citep{owe95} corresponds to the luminosity distance of $d_{\rm L} = 391$\,Mpc and the conversion scale of 1.604\,kpc per\,arcsecond.

\section{Radio Data}
\label{S-radio}

Archival VLA data at two frequencies were analyzed to compare the 4C\,+67.13 radio morphology at arcsecond-resolution with the newly acquired {\it Chandra} X-ray maps. We combined multi-configuration 1.5\,GHz data obtained in 1991 when the VLA was in its A- (Sep 8; program AL238, 6 min.\ integration) and C-arrays (Jan 29; AO104, 5.5 min.), together with B-array data from 2001 May 7 (AC582, 1.5 hrs). At 4.7\,GHz, we analyzed a single deep (2 hr) C-array dataset from 2001 July 20 (program AC582). We used AIPS for the standard calibration and DIFMAP \citep{she94} for self-calibration and imaging. The resultant map beamsizes were $3.12\arcsec \times 2.03\arcsec$ (position angle, $PA = 43.2^\circ$) at 1.5\,GHz and $5.05\arcsec \times 3.21\arcsec$ ($PA=12.4^\circ$) at 4.7\,GHz.

The VLA contours of 4C\,+67.13 superimposed on the Sloan Digital Sky Survey \citep[SDSS;][]{yor00} optical image are shown in Figure\,\ref{VLAimage} (left panel). Note that despite its edge-brightened FR\,II radio morphology, the studied radio galaxy does not obviously show compact high surface brightness hotspots at the edges of the radio lobes. With a total linear size of $47\arcsec \simeq 75$\,kpc and axial ratio $\simeq 2$, which is typical for FR\,IIs, the entire volume occupied by the radio lobes is $V_{\ell} \simeq 1.6 \times 10^{69}$\,cm. The radio core of 4C\,+67.13 coincides with the BCG, which is the brighter member of the galaxy pair CGPG\,0719.8+6704, and which is offset by $\simeq 38\arcsec \simeq 61$\,kpc from the position of the cluster center (as inferred from the modeling of the cluster X-ray surface brightness distribution; see \S\,\ref{sec:2dfit} below). 

The radio core of 4C\,+67.13 is distinguished by its inverted radio spectrum, with 1.5\,GHz peak of 8.4\,mJy beam$^{-1}$ and 4.7\,GHz peak of 12.1\,mJy beam$^{-1}$. Subtracting the estimated contribution of the core flux from the lobes, the 1.5\,GHz fluxes are 0.314\,Jy (eastern) and 0.331\,Jy (western), and the 4.7\,GHz fluxes are 0.120\,Jy (eastern) and 0.127\,Jy (western). The resultant spectral indices for both lobes are $0.84 \pm 0.12$, assuming $10\%$ uncertainties in each flux measurement, and the monochromatic lobes' radio power reads as $L_{1.5\,{\rm GHz}} \simeq 1.8 \times 10^{41}$\,erg\,s$^{-1}$. The total fluxes and the lobe spectra from the VLA maps compare well with those derived from integrated measurements at 1.4\,GHz \citep[NVSS][]{con98} and 4.85\,GHz \citep{gre91} as well as the spectral index $\alpha = 0.88 \pm 0.08$ measured using 2.7, 5, and 10.7 GHz Effelsberg data \citep{rei00}. At lower frequencies, we derived $\alpha = 0.64 \pm 0.04$ by combining integrated fluxes at 74\,MHz \citep{lan12}, 178\,MHz \citep{gow67}, and 365\,MHz \citep[][assuming $15\%$ uncertainty]{dou96}, together with the NVSS measurement, indicating flatter spectra below $\sim1$ GHz.

\begin{figure}[tbp]
\begin{center}
\includegraphics[width=0.9\columnwidth]{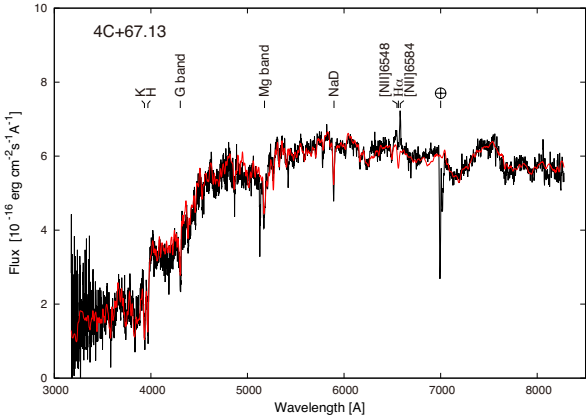}
\caption{The rest frame optical spectrum of 4C\,+67.13, obtained with WHT. The observed spectrum is plotted in black while the STARLIGHT model is drawn in red. Unremoved telluric lines are marked with $\oplus$.}
\label{opt_spec}
\end{center}
\end{figure}

The right panel in Figure\,\ref{VLAimage} presents the VLA contours of the WAT radio galaxy located in the southern part of the Abell\,578 cluster. The distance between the radio core of this source and the position of the cluster X-ray centroid is about $158\arcsec$, corresponding to 253\,kpc.

Although 4C\,+67.13 lacks clearly defined hotspots, and its radio luminosity is intermediate between FR\,I and II sources according to the often considered luminosity divide, it should nonetheless be classified as an FR\,II radio galaxy. The original classification by \citet{fan74} is based on the ratio of the distance between the regions of the highest brightness on opposite sides of the radio core, to the total extent of the radio structure; sources with this ratio greater than 0.5 are classified as FR\,IIs (``edge-brightened radio morphology''). In the case of 4C\,+67.13, we obtain the ratio of $\simeq 0.65$ for the higher resolution map from \citet{owe97}, and $0.58$ for the lower resolution map (Figure\,\ref{VLAimage}, left). The relatively low radio luminosity of the analyzed object compared with `classical' FR\,IIs should not be considered as a surprise, as recent studies reveal the existence of a large number of FR\,IIs near and below the classical luminosity divide \citep{bes09,gen13}. An alternative interpretation could be that the edge-brightened appearance of the radio lobes in 4C\,+67.13 is in fact only due to a special alignment of a `narrow-angle tailed', i.e. intrinsically edge-dimmed radio structure viewed with its prolonged tails along our line of sight.
That 4C +67.13 was selected for study from a parent sample \citep{owe97} where the BCGs predominantly show clear FR I morphology, would indicate an increase in the probability for such a special alignment.
Although we think it is improbable that both tails would be aligned in such a way that their extended diffuse radio emission is undetectable, and the resulting structure --- characterized notably by sharp boundaries --- appears highly symmetric in projection, we can not completely rule out this scenario.

\begin{figure}[tbp]
\vspace{-8mm}
\begin{center}
\includegraphics[width=0.6\columnwidth,angle=-90]{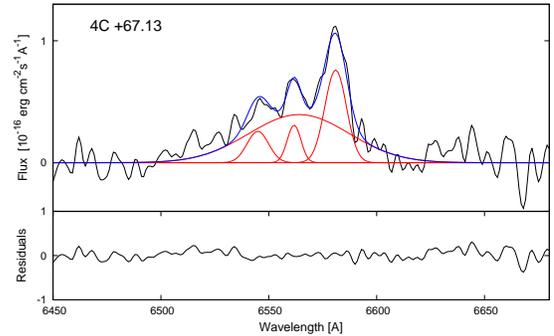}
\caption{Observed [NII]$\lambda$6548,6584 + H$\alpha$ blended emission lines of 4C\,+67.13 (plotted in black), with Gaussian profiles fitted to the emission lines overlaid in red. The reconstructed blend is drawn in blue in the top panel while the residuals are shown in the bottom panel.}
\label{opt_spec2}
\end{center}
\end{figure}

\section{Optical Spectroscopy}
\label{S-opt}

Optical spectroscopic observations of the BCG hosting 4C\,+67.13 and its neighboring galaxy were obtained with the WHT on 2010 April 21. Longslit spectra were taken using the Intermediate dispersion Spectrograph and Imaging System (ISIS) double-armed spectrograph with a chosen slit width of 1\arcsec\ (corresponding to 1.6\,kpc). The slit was centered on the nucleus of 4C\,+67.13 and was oriented along the position angle of the galaxy pair ($PA=93\deg$) so data were obtained for both objects. To identify and reject cosmic rays, the integration time was split into two 900\,s exposures per arm. The average seeing FWHM during the observing run was about 0.8\arcsec.

In the blue arm, the grism R300B was used and the detector was a thinned, blue-sensitive EEV12, array of $4096\times2048$ (13.5 micron) pixels with a spatial scale of 0.20\arcsec\,pixel$^{-1}$. The binning factor was $1\times1$ which yielded a wavelength coverage between about 3300\,{\AA} and 5300\,{\AA} with a dispersion of 0.86\,{\AA}\,pixel$^{-1}$. The instrumental resolution was 3\,{\AA} (FWHM) corresponding to $\sigma_{\rm inst}\sim90$\,km\,s$^{-1}$. In the red arm, the grism R158R with an order blocking filter GG495 was used. The detector was the default chip for the ISIS red arm, RED+, a red-sensitive array of $4096\times2048$ (15.0 micron) pixels with spatial scale 0.224\arcsec\,pixel$^{-1}$. The binning factor was $1\times1$ which yielded a wavelength coverage between about 5300\,{\AA} and 9000\,{\AA} with a dispersion of 1.82\,{\AA}\,pixel$^{-1}$. The instrumental resolution was 6\,{\AA} corresponding to $\sigma_{\rm inst}\sim102$\,km\,s$^{-1}$.

\begin{table}[bp]
\caption{Optical emission lines in the 4C\,+67.13 nucleus}
\begin{center}
\begin{tabular}{cccc}
\hline\hline
Line	& Flux & Position\footnotemark[1] & FWHM\\
	& [$10^{-15}$\,erg\,cm$^{-2}$\,s$^{-1}$] & [\AA] & [km\,s$^{-1}$]\\
\hline
[NII]$\lambda 6548$ 			& 0.30 & 6545.6 & 485.7\\
H$\alpha_\mathrm{narrow}$ 	& 0.32 & 6561.7 & 484.7\\
H$\alpha_\mathrm{broad}$ 	& 2.12 & 6567.9 & 2336.9\\
$\mathrm{[NII]}$$\lambda 6584$ 			&0.895& 6581.0 & 485.7\\
\hline
\end{tabular}
\footnotetext[1]{The redshift of $z=0.087227$ \citep{gom97} is adopted.}
\end{center}
\label{opt_flux}
\end{table}

Reduction steps were performed for both spectral ranges separately using the NOAO IRAF packages. A master bias frame was created by averaging all the bias frames obtained during the observing night and subtracted from the science frames. Also, a master flat field frame was created and all 2D science frames were corrected for flat field, then cosmic rays were removed in combining the science exposures. Accurate wavelength calibration was performed using ArNe lamp exposures obtained before and after every target exposure and checked using sky lines, and the correction for optical distortion was applied. The contribution from the sky was determined from the sky regions at the two sides of the resulting spectrum, and then subtracted. The extraction of the 1D spectra was performed using APALL task. Scientific exposures were flux calibrated using the standard SP1257+038. The slit position angle falls along the radio jet/lobe axis, which is perpendicular to the optical photometric major axis so light from a region of $14.3\arcsec \times 1\arcsec$ (23\,kpc\,$\times$\,1.6\,kpc) was summed into the 1D spectrum.

\begin{figure}[tbp]
\begin{center}
\includegraphics[width=\columnwidth]{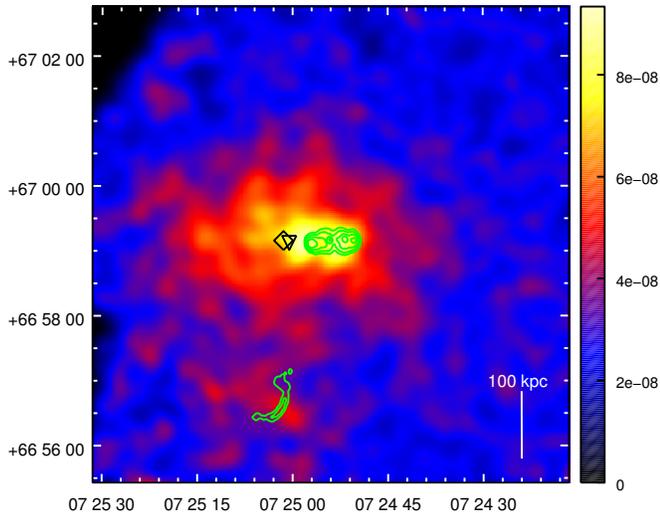}
\caption{The exposure-corrected {\it Chandra} 0.5--7.0\,keV image of Abell\,578 with the VLA 4.7\,GHz contours superimposed.  X-ray point sources, including the nuclei of the two cluster radio galaxies, are removed from the image. The pixel size is equal to 4 times of the ACIS pixel size (i.e., $4\times 0.492\arcsec$ = 1.968\arcsec) and the image is smoothed by a gaussian with $\sigma=16\times$ the $\mathrm{ACIS~pixel~size}$. Open black triangle and square denote the position of the cluster center obtained from the 2-dimensional fitting with elliptical and isotropic 2-D $\beta$ models, respectively (see \S\,\ref{sec:2dfit}.) }
\label{image}
\end{center}
\end{figure}

Before fitting the stellar continuum, the 1D spectra were corrected for Galactic extinction using the \citet{sch98} maps and the extinction law of \citet{car89}, shifted to the rest frame using a redshift of $z=0.087227$ \citep{gom97} and resampled to $\Delta\lambda=1$\,\AA. Analysis of the galaxy stellar continuum was performed using the STARLIGHT code \citep{cid05} which fits an observed spectrum with a linear combination of the simple stellar populations (SSPs) extracted from the models of \citet{bru03}. Bad pixels and emission lines were masked and left out of the fits. Two bases of SSPs used in these fits: (1) $N_*=150$ with 25 ages between 1\,Myr and 18\,Gyr and six metallicities from $Z= 0.005 \, Z_{\odot}$ to $2.5 \, Z_{\odot}$, and (2) $N_*=45$ with 15 ages between 1\,Myr and 13\,Gyr and three metallicities from 0.004 to $2.5 Z_\odot$. The results of the stellar populations and velocity dispersions obtained from the fits made with these two bases were consistent.

Stellar population fits to the observed spectrum of the BCG (Figure \ref{opt_spec}) revealed only old populations with 11 and 13\,Gyr and metallicity of $0.02 Z_\odot$; no traces of young populations (with ages below a few Gyrs), indicative of a recently enhanced starformation, have been found in the modelling. The velocity dispersion obtained from the fits was $\sigma_{\star} = 321 \pm 17$\,km\,s$^{-1}$ which gives a SMBH mass of $\log M_{\rm BH}/M_\odot=8.96\pm0.22$ based on the $M_{\rm BH} - \sigma_{\star}$ relation from \citet{tre02}. Taking instead the $M_{\rm BH} - \sigma_{\star}$ relation from \citet{fer05}, one obtains $\log M_{\rm BH}/M_\odot = 9.21 \pm 0.27$.

\begin{table}[tbp]
\caption{Results of the 2-dimensional fitting to the X-ray image}
\begin{center}
\begin{tabular}{lcc}
\hline\hline
Parameter & \multicolumn{2}{c}{Value}\\
\hline
\multicolumn{3}{c}{Cluster region}\\
& isotropic & elliptical\\
\hline
Core radius $r_C$ [kpc] & $118^{+43}_{-29}$ & $170^{+56}_{-38}$\\
Center position $\alpha$\footnotemark[1] & $7^\mathrm{h}25^\mathrm{m}01^\mathrm{s}.4^{+0.6}_{-0.6}$ & $7^\mathrm{h}25^\mathrm{m}00^\mathrm{s}.6^{+0.6}_{-0.6}$ \\
Center position $\delta$\footnotemark[1] & $+66^\circ 59\arcmin 10\arcsec ^{+2}_{-3}$ & $+66^\circ 59\arcmin 09\arcsec ^{+2}_{-2}$ \\
Ellipticity $\epsilon$ & --- & $0.32^{+0.04}_{-0.05}$\\
Ellipticity $\theta$ [degree] & --- & $-3.2^{+4.3}_{-4.2}$\\
Amplitude\footnotemark[2] [$10^{-9}$] & $2.83^{+0.30}_{-0.25}$ & $2.96^{+0.31}_{-0.26}$ \\
Index $\beta$ & $0.58^{+0.17}_{-0.10}$ & $0.69^{+0.23}_{-0.12}$ \\
reduced {\it C}-stat/dof & 1.17/600 & 1.11/598 \\
\hline
\multicolumn{3}{c}{Background region}\\
\hline
Constant\footnotemark[2] [$10^{-9}$] & \multicolumn{2}{c}{$1.54^{+0.02}_{-0.02}$} \\
reduced {\it C}-stat/dof & \multicolumn{2}{c}{1.15/1950} \\
\hline
\end{tabular}
\footnotetext[1]{J2000 equinox}
\footnotetext[2]{$\mathrm{photon~pixel^{-2}~cm^{-2}~s^{-1}}$}
\end{center}
\label{2dfit_param}
\end{table}

The [NII]$\lambda$6548,6584 and H$\alpha$ lines in the stellar light subtracted spectra were fitted simultaneously with Gaussian profiles. Fitting was performed using the SPECFIT task in the STSDAS external IRAF package. Lines of the same ion were assumed to have the same offset and width and the additional constraint [NII]$\lambda$6584/[NII]$\lambda6548=3$ was further imposed on the line flux ratio. The results are presented in Figure\,\ref{opt_spec2} and Table\,\ref{opt_flux} (where the provided line intensities are not extinction corrected). Detailed optical activity classification using line diagnostic diagrams \citep{kew06,but10} is not possible in the case of 4C\,+67.13 because only [NII] and H$\alpha$ emission lines were detected with S/N ratio larger than 3.
Nevertheless, these lines can still be used for a spectral diagnostic following \citet{cid11}, indicating that the 4C\,+67.13 nucleus is of a LINER type (`low-ionization nuclear emission-line region').
The H$\alpha$ luminosity $L_{\rm H\alpha} = 5.9 \times 10^{39}$\,erg\,s$^{-1}$ translates to the bolometric accretion-related luminosity $L_{\rm nuc} \simeq 10^{43}$\,erg\,s$^{-1}$ \citep{sik13}, or in the Eddington units $\Lambda \equiv L_{\rm nuc}/L_{\rm Edd} \simeq 10^{-4}$. This low accretion rate is confirmed by the analysis of the {\it Chandra} data for the 4C\,+67.13 nucleus presented below in \S\,\ref{S-nucleus}.

\begin{figure*}[tbp]
\begin{center}
\includegraphics[width=\columnwidth]{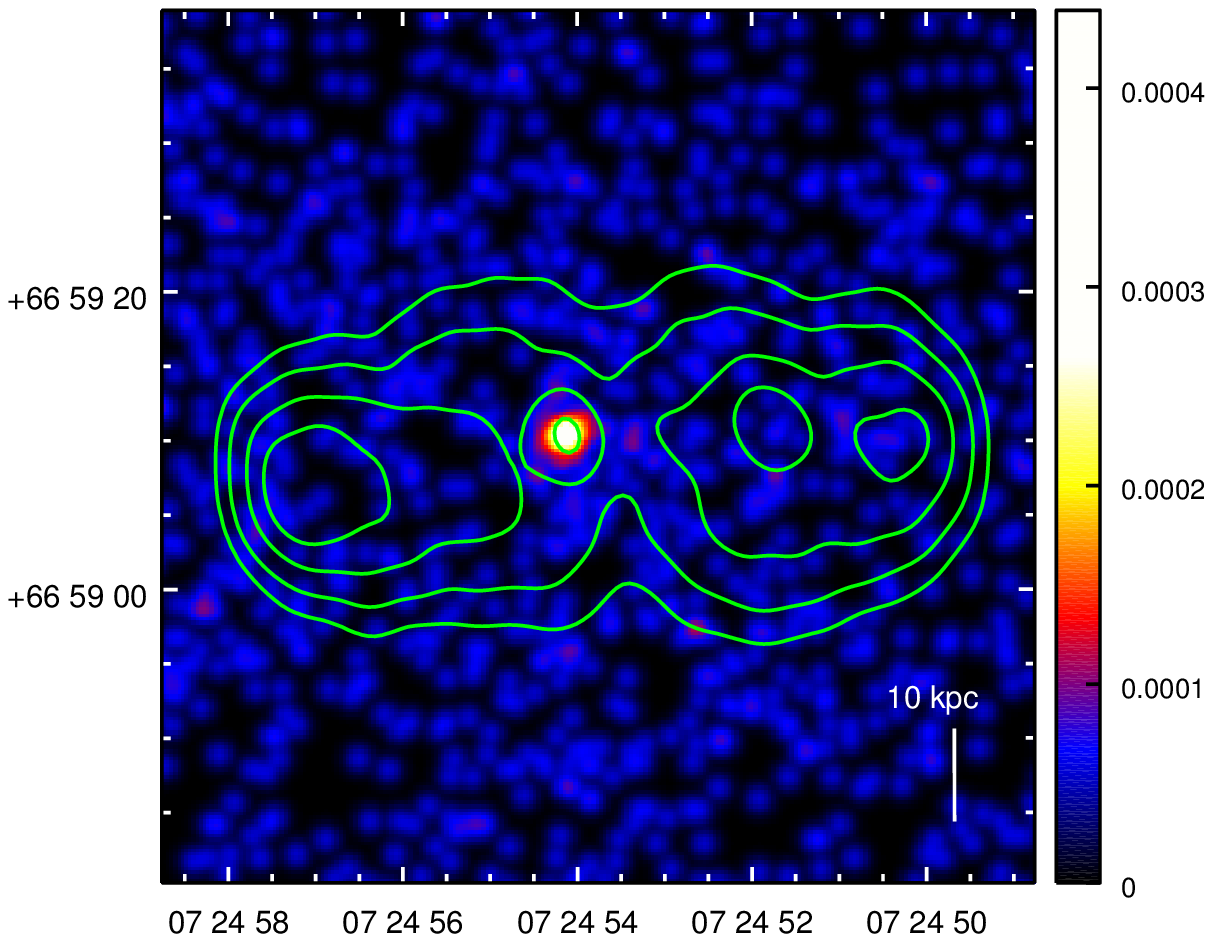}
\hspace{3.0mm}
\includegraphics[width=\columnwidth]{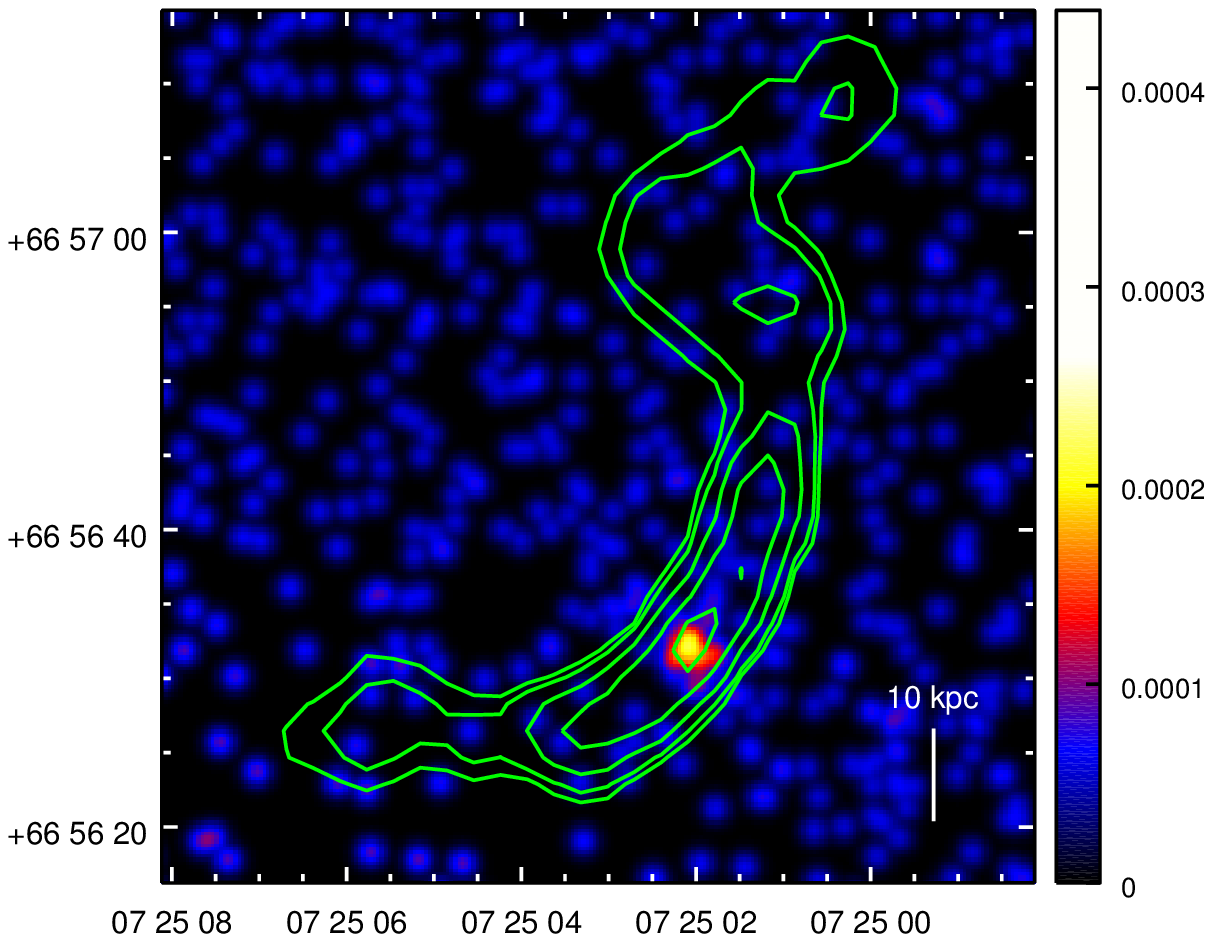}
\caption{
{\it Chandra} zoom-in images of the central part of the Abell\,578 cluster around the position of 4C\,+67.13 (left), and of the southern part of the cluster around the position of the WAT radio galaxy (right), with the VLA 4.7\,GHz radio contours superimposed. X-ray point sources are not removed from these figures. The pixel size is equal to a half of the ACIS pixel size (i.e., $0.5\times 0.492\arcsec$). The image is smoothed by a Gaussian with $\sigma=\mathrm{ACIS~pixel~size}$.
}
\label{image-zoom}
\end{center}
\end{figure*}

The analogous analysis performed for the second, fainter galaxy of the CGPG 0719.8+6704 pair revealed $\sigma_{\star} = 233 \pm 97$\,km\,s$^{-1}$, $\log M_{\rm BH}/M_\odot \simeq 8.3-8.7$,
old stellar populations with a small addition of a younger one (1.43\,Gyr, $Z= 0.05\,Z_{\odot}$), and a LINER-type nucleus with $L_{H\!\alpha} \simeq 10^{40}$\,erg\,s$^{-1}$ or $\Lambda \simeq 5 \times 10^{-4}$.

\section{Chandra X-ray Observations}

Abell\,578 was observed with {\it Chandra} on 2010 May 29 (obsid$=$11749) and July 23 (obsid$=$12225) using the ACIS-S detector. The source was placed on the back-illuminated CCD (S3) and the observations were made in very faint (VFAINT) mode with a total exposure time of $39.3$\,ks.

Analysis was performed with the CIAO version 4.5 software. We processed the data by running CIAO tool {\sf chandra\_repro} and applied the calibration files from CALDB 4.5.5.1. In order to obtain the highest angular resolution image, the sub-pixel algorithm EDSER was also applied. Event files of the two observations were merged into a single event file with {\sf reproject\_obs} for the image analysis. Point sources detected with {\sf wavdetect} were removed, including also the nucleus of 4C\,+67.13, whose spectrum is analyzed separately in \S\,\ref{S-nucleus}. Besides the 4C\,+67.13 nuclear source, there were 15 X-ray sources found in the cluster region ($<2.5\arcmin$ of the cluster center obtained in \S\,\ref{sec:2dfit}) and 30 sources in the analyzed region. All events corresponding to the detected point sources are excluded from the analysis of the extended emission. Spectral extraction was performed with {\sf specextract} and modeling was done in Sherpa \citep{fre01,ref09}. The spectra and response files were extracted from the individual data and merged with {\sf combine\_spectra}. All spectral models were fitted to the data in the 0.5--7.0\,keV energy range. Except for the radial profile analysis, we used the {\it C}-stat fitting statistics \citep{cas79} and the Nelder-Mead optimization method \citep{nel65}.

\subsection{Image Analysis}

Figure\,\ref{image} shows the exposure-corrected and smoothed {\it Chandra} image of Abell\,578 in the 0.5--7.0\,keV energy band. Events of the two separate observations were summed to create the image. The green contours are from the VLA 4.7\,GHz radio map discussed in \S\,\ref{S-radio} above. 

Figure\,\ref{image-zoom} presents the {\it Chandra} zoom-in images of the central and southern parts of the cluster, around the positions of the two cluster radio galaxies (left and right panels, respectively). Point sources were not removed from these figures. No obvious small-scale X-ray sub-structures adjacent to the radio lobes can be noted, most likely due to the limited photon statistics.

\subsubsection{2-Dimensional Fitting}
\label{sec:2dfit}

\begin{figure*}[tbp]
\begin{center}
\vspace{-8.0mm}
\includegraphics[width=13cm]{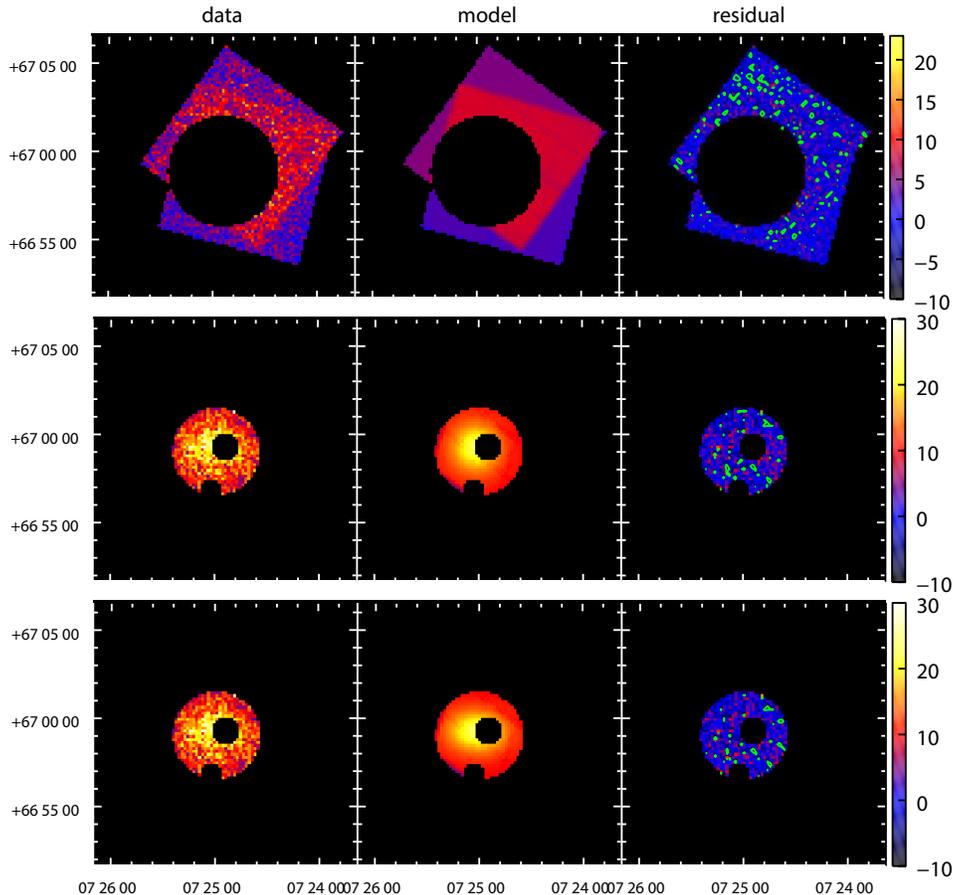}
\caption{Two-dimensional analysis of the background (upper panel) and the Abell\,578 cluster region (middle and lower panels). The ellipticity parameters are fixed to the circular symmetric values in the middle panel, while they are free parameters in the lower panel. The images of the data (left), adopted  {\sf beta2d+const2d} model (center), and residual (right) are shown in the same color scale. The green contours overlaid on the residual images represent $\pm$2-sigma of the error of the data count in each pixel. The black regions in the images are ignored in the fitting. The pixel size is $9.84\arcsec$ (20 times of the original ACIS pixel size).}
\label{2dfit}
\end{center}
\end{figure*}

To estimate the position of the cluster center and to extract the general characteristics of the cluster large-scale morphology, two-dimensional fits to the {\it Chandra} data was performed. Figure\,\ref{2dfit} shows the corresponding images of the background regions (upper panel) and of the cluster (middle and lower panels). Total counts are 11720 for the background and 8058 for the source.

First, we modeled the background image with the constant model ({\sf const2d}) and the source image with the isotropic 2-D $\beta$ model ({\sf beta2d}). The regions around the two radio sources were removed from the analysis; these correspond to the circular black spots in the cluster images in Figure\,\ref{2dfit}. In the fitting, an exposure map created by running {\sf flux\_obs} was used to correct non-uniformity of the effective area in the ACIS chip. The background model was fixed when fitting the source region. The fitting results are listed in Table\,\ref{2dfit_param}. In particular, we found the position of the cluster center $(\alpha, \delta)=(7^\mathrm{h}25^\mathrm{m}01^\mathrm{s}.4^{+0.6}_{-0.6}, +66^\circ 59\arcmin 10\arcsec ^{+2}_{-3})$, the cluster core radius $r_C = 118^{+43}_{-29}$\,kpc, and the index $\beta=0.58^{+0.17}_{-0.10}$ (reduced fit $C$-statistic 1.17/600), consistent with the values obtained by \citet{gom97}.

We also modeled the source image with the elliptical 2-D $\beta$ model. The results are given in the third column of Table\,\ref{2dfit_param}. The thus evaluated position of the cluster center is consistent with the result of the isotropic $\beta$ model, and the ellipticity parameters are $\epsilon=0.32^{+0.04}_{-0.05}$ and $\theta=-3.2^{+4.3}_{-4.2}$\,deg, while the other parameters changed only slightly ($r_C = 170^{+56}_{-38}$\,kpc, $\beta=0.69^{+0.23}_{-0.12}$). The elliptical model gives, however, a better fit with the reduced fit $C$-statistic of 1.11/598. In this way, we confirmed and robustly quantified the E-W elongation of the cluster mentioned by \cite{gom97}.

The positions of the cluster center obtained from our analysis are shown in Figure\,\ref{image} where the black triangle and square denote the values obtained with the elliptical and isotropic models, respectively. The cluster center position is clearly offset from the position of the 4C\,+67.13 core. Our radio data indicate that the position of the 4C\,+67.13 core is $(7^\mathrm{h}24^\mathrm{m}54^\mathrm{s}.150, +66^\circ 59\arcmin 10.30\arcsec)$. Therefore, the offset in projection is $r_x \simeq 38\arcsec$, corresponding to 61\,kpc for the elliptical cluster model.

\subsubsection{Radial Profiles}
\label{S-radial}

\begin{figure}[tbp]
\begin{center}
\vspace{-3.0mm}
\includegraphics[width=\columnwidth]{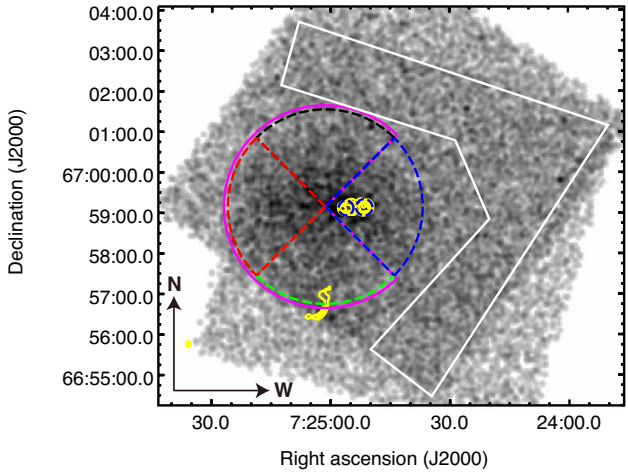}
\caption{{\it Chandra} X-ray image of Abell\,578 (gray scale) overlaid with the VLA 4.7\,GHz contours (yellow). Different regions selected for constructing radial surface brightness profiles of the cluster are denoted by magenta, red, black, green, and blue contours (see \S\,\ref{S-radial}). Region defined by a white polygon was used to extract the background surface brightness (``background A").}
\label{reg_profile}
\end{center}
\end{figure}

\begin{figure}[tbp]
\begin{center}
\includegraphics[width=\columnwidth]{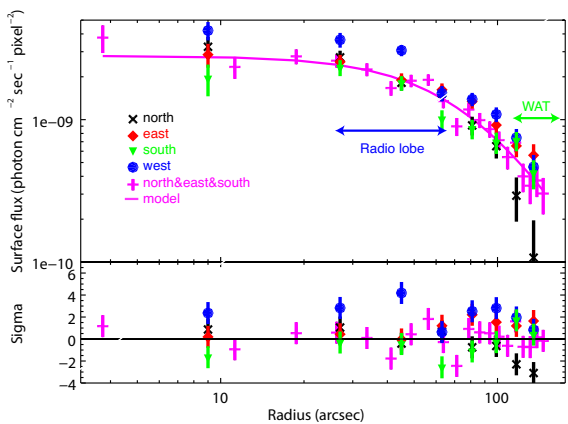}
\caption{One-dimensional radial X-ray surface flux profiles of the Abell\,578 cluster extracted from different regions defined in \S\,\ref{S-radial} (color-coded symbols denote the corresponding regions in Figure\,\ref{reg_profile}). Thick magenta curve represents the best-fit {\sf beta1d+const1d} model applied to the region enclosed by the magenta solid line in Figure\,\ref{reg_profile}.}
\label{fig_rad}
\end{center}
\end{figure}

A radial surface brightness profile of the central parts of the Abell\,578 cluster was constructed from a pie region enclosed by the magenta solid line shown in Figure\,\ref{reg_profile}. The center of the pie region was set at the position of the cluster center obtained from the two-dimensional fitting discussed in the previous section. We extracted surface brightness from 20 annuli covering a radius of $150\arcsec$ ($\simeq241$\,kpc) by utilizing the {\sf dmextract} tool. A pie region embedding within the 4C\,+67.13 radio source was ignored in order to exclude the expected contribution from the cluster gas interacting with the expanding radio lobes, as well as the X-ray emission of the radio galaxy itself (including the non-thermal emission of the extended lobes as well as the radiative output of the AGN and circumnuclear gas). We subtracted the surface brightness of the region defined by a polygon as a background. Hereafter we refer to this background region as ``background A.'' We fitted the background-subtracted surface flux profile with the {\sf beta1d} model using the {\sf chi2datavar} statistics. The fitting results are presented as the magenta curve in Figure\,\ref{fig_rad}, and the best-fit parameters are listed in Table\,\ref{par_rad}.

\begin{table}[bp]
\caption{Results of the X-ray radial profile analysis}
\begin{center}
\begin{tabular}{lcc}
\hline\hline
Parameter & Value\\
\hline
Core radius $r_C$ [kpc] &  $138^{+49}_{-32}$\\
Amplitude\footnotemark[1] [$10^{-9}$] & $2.80^{+0.23}_{-0.20}$ \\
Index $\beta$ & $0.71^{+0.26}_{-0.14}$ \\
$\chi_\nu^2$/dof & 1.25/17 \\
\hline
\end{tabular}
\footnotetext[1]{$\mathrm{photon~pixel^{-2}~cm^{-2}~s^{-1}}$}
\end{center}
\label{par_rad}
\end{table}

In order to investigate the impact of the central radio source on the surrounding cluster gas and to further examine the cluster elongation, we also constructed radial surface brightness profiles separately for the north, east, south, and west parts of the cluster center. The respective black, red, green and blue dashed regions in Figure\,\ref{reg_profile} were used to construct these profiles. Each region was divided into 8 annuli, with the radio lobes and X-ray point sources removed. The profiles cover the regions from the center to a radius of $144\arcsec$ ($\simeq 231$\,kpc) within an angle of $90^\circ$. The results are shown in Figure\,\ref{fig_rad}. The cluster surface brightness in the western region (blue filled circles), where the central radio source 4C\,+67.13 is located, displays a clear excess over the average model. This may be due to the gas compression/heating by the expanding radio lobes. Also, for radii $\gtrsim 80$\arcsec, the surface brightness in the northern region (black crosses) is slightly lower than the average model, while that in the eastern region (red diamonds) slightly higher, consistent with the E-W elongation of the cluster atmosphere.

The last two data points in Figure\,\ref{fig_rad} for the southern region (green triangles) overlap with the position of the WAT radio galaxy. While we expect the elongation of the cluster to reduce the surface brightness along this direction compared to the average model, this effect is compensated for by the cluster substructure discussed also in \citet{gom97}, therefore no surface brightness decrement is seen.

\subsection{Spectral Analysis}

\subsubsection{Cluster Gas Around 4C\,+67.13}
\label{sec:specfit}

\begin{figure}[tbp]
\begin{center}
\vspace{-3.00mm}
\includegraphics[width=\columnwidth]{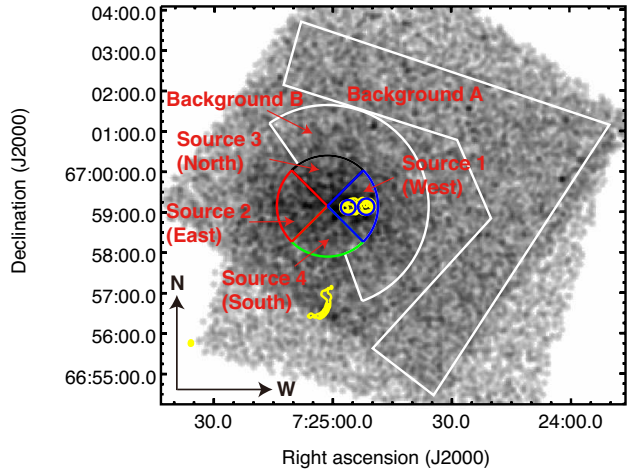}
\caption{{\it Chandra} X-ray image of Abell\,578 (gray scale) overlaid with the VLA 4.7\,GHz contours (yellow). Different regions selected for the cluster spectral analysis are denoted by red, black, green, and blue contours (see \S\,\ref{sec:specfit}). Regions defined by white polygons were used to extract the background spectra (``background A" and ``background B").}
\label{reg_spec}
\end{center}
\end{figure}

The analysis presented in the previous section provided some indication for the interaction of the 4C\,+67.13 with the surrounding gas. In order to investigate in more detail the impact of the radio source on the cluster environment, we extracted spectra from four smaller pie regions around the cluster center, with the outer radius of $75\arcsec$, as shown in Figure\,\ref{reg_spec}. We refer to the western, eastern, northern, and southern regions as ``source 1", ``source 2", ``source 3", and ``source 4", respectively. From source\,1, the emission from the positions of the radio lobes and the core of 4C\,+67.13 were once again removed. For the background, we chose the polygon region A, the same as the  one used in constructing the cluster radial profiles (see \S\,\ref{S-radial}), and a large annular region ``background B.'' While background A represents a uniform background in the ACIS chip --- i.e., consisting of the Cosmic X-ray Background (CXB), Galactic Halo (GH) emission, and the instrumental background --- background B was chosen to characterize the emission from the outer parts of the cluster adjacent to the selected source regions 1--4.

First we fitted the spectra of backgrounds A and B (total counts in the 0.5--7.0\,keV energy band $3464$ and $3487$, respectively). For the instrument background, we used the model defined in ``{\sf acis\_bkg\_model.py}"\footnote{
\url{https://github.com/taldcroft/datastack/blob/master/acis_bkg_model.py}
}
that consists of two power-laws and six gaussians.  Parameters in the model were fixed at the values for {\sf ccd\_id = 7}, and only the normalization was a free parameter. We used a constant ARF by setting all values in {\sf specresp} column to 100. The CXB and GH components were modeled by an absorbed power-law model and an unabsorbed thermal plasma model APEC, respectively. The photon index of the CXB was fixed at 1.41 \citep{del04} and the temperature of the GH was fixed at $0.2$\,keV \citep{kun00}. The absorption column density for the CXB was fixed at the Galactic value, $4.39\times 10^{20}$\,cm$^{-2}$. The spectrum of background B was modeled by the background A scaled for the extraction region and an APEC model with Galactic absorption. The redshift of the APEC model for background B was fixed at 0.0866, and the abundance was allowed to vary. The obtained best-fit parameters of backgrounds A and B are listed in Table\,\ref{par_spec_bkg}. The resulting normalization of the CXB component $12.4\pm1.5$\,ph\,keV$^{-1}$\,cm$^{-2}$\,s$^{-1}$\,str$^{-1}$ at 1\,keV is consistent with that of \citet{del04}. These background model parameters were then fixed in the following analysis.

In the above, we have assumed that background A does not contain any significant cluster emission. We verified this assumption by adding an APEC component with Galactic absorption to the model for background A and obtained an upper limit for the APEC normalization, with the 90\% confidence intervals of $7.6\times10^{-6}$ for 2\,keV and $2.4\times10^{-6}$ for 1\,keV, where the temperature of the APEC component is fixed. Since the backscale value of source 1 is $\sim0.1$, this upper limit corresponds therefore to a negligible amount of the APEC component for source 1 spectrum.

Next we used the CIAO Sherpa extension package {\sf deproject}\footnote{\texttt{http://cxc.harvard.edu/contrib/deproject/}} \citep{sie2010} for spectral analysis of the selected source regions. Here, we assumed spherical symmetry for the cluster gas. In the analysis, the spectral model was composed of a linear weighted combination of a source model, the background B model, and the background A model was scaled by the backscale value: {\sf rsp(W$\mathsf{_{src}}$*src\_model+W$\mathsf{_{bkg}}$*bkgB\_model) + bkg\_scale*(bkgA\_rsp(bkgA\_model))}. The weight was defined as a volume of extraction region normalized by a volume of a sphere enclosing background B. The weight was calculated in the {\sf deproject} tool except for the source 1, for which the weight was calculated by Monte-Carlo integration. The weight of the source 1 region is $0.0245852\pm0.0000008$ and that of background B for the source 1 analysis is $0.0462102\pm0.0000009$. We assumed an APEC model with Galactic absorption for each spectrum. We fixed the abundance at 0.0, 0.5 and 1.0 at first, and then set it free.

\begin{table}[tbp]
\caption{Best-fit model parameters for backgrounds A and B}
\begin{center}
\begin{tabular}{llc}
\hline\hline
Model component & Parameter & Value\\
\hline
\multicolumn{3}{c}{background A}\\
\hline
Instrumental background & normalization & $31.42^{+0.73}_{-0.72}$\\
Cosmic X-ray background & photon index & $1.41$ (fix)\\
& norm\footnotemark[1] [$10^{-6}$] & $23.7^{+2.9}_{-2.8}$\\
Galactic Halo & $kT$ [keV] & 0.2 (fix)\\
& norm\footnotemark[2] [$10^{-6}$] & $5.3^{+3.2}_{-3.1}$\\
Fit statistics & reduced {\it C}-stat/dof & $0.64 / 60$\\
\hline
\multicolumn{3}{c}{background B}\\
\hline
Cluster gas emission & Abundance & $0.34^{+0.23}_{-0.15}$\\
& $kT$ [keV] & $2.50^{+0.43}_{-0.37}$\\
& norm\footnotemark[2] [$10^{-3}$] & $0.56 ^{+0.07}_{-0.07}$\\
& density [$10^{-3}~\mathrm{cm^{-3}}$] & $0.78 ^{+0.05}_{-0.05}$\\
Fit statistics & reduced {\it C}-stat/dof & $1.04 / 51$\\
\hline
\end{tabular}
\footnotetext[1]{Normalization of the power-law model in units of $\mathrm{photon~cm^{-2}~s^{-1}}$ at 1 keV.}
\footnotetext[2]{Normalization of the APEC model defined as $10^{-14}/(4\pi\left[D_A(1+z)\right]^2)\int n_en_H dV$.}
\end{center}
\label{par_spec_bkg}
\end{table}

\begin{figure*}[tbp]
\begin{center}
\vspace{1mm}
\includegraphics[width=\columnwidth]{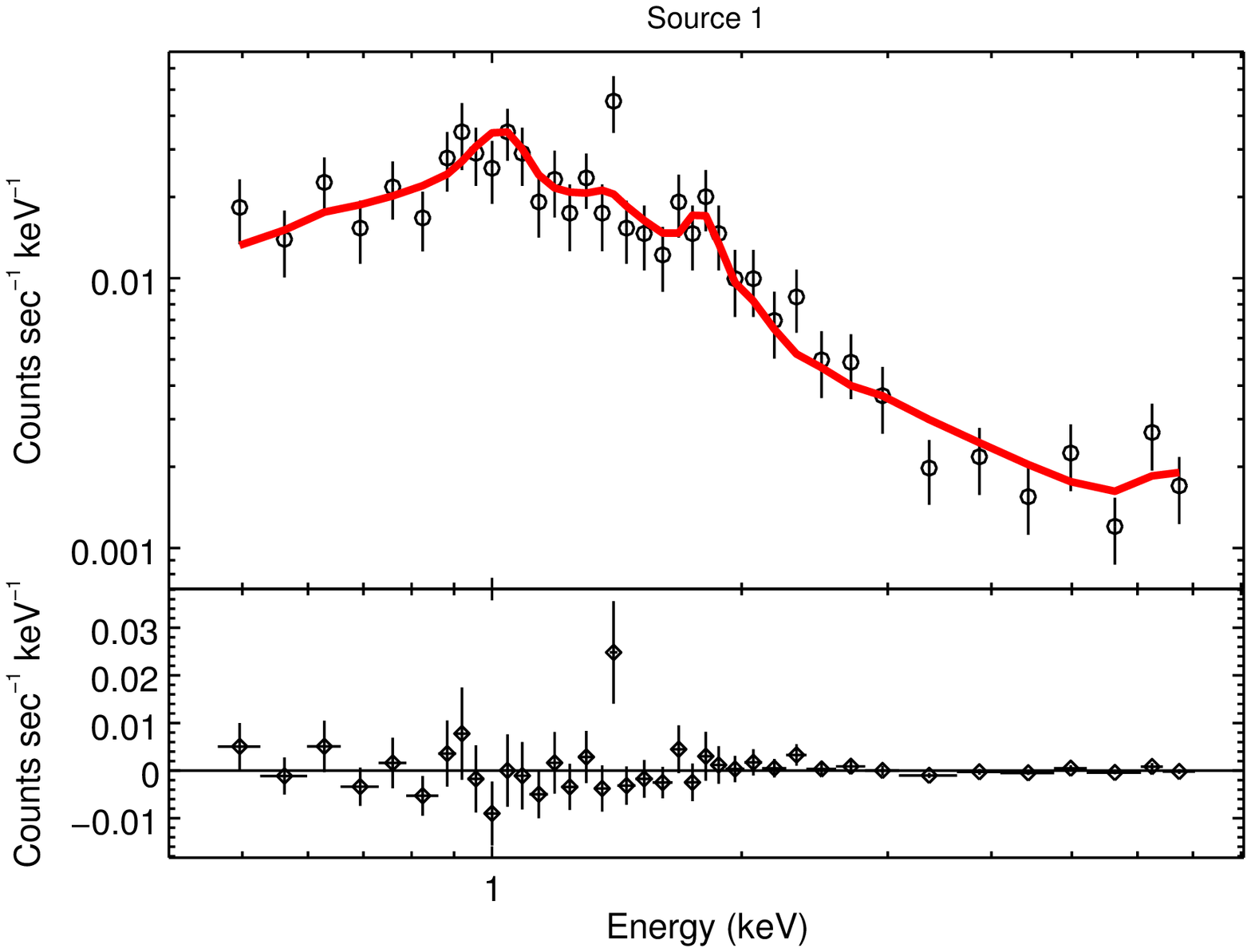}
\hspace{3mm}
\includegraphics[width=\columnwidth]{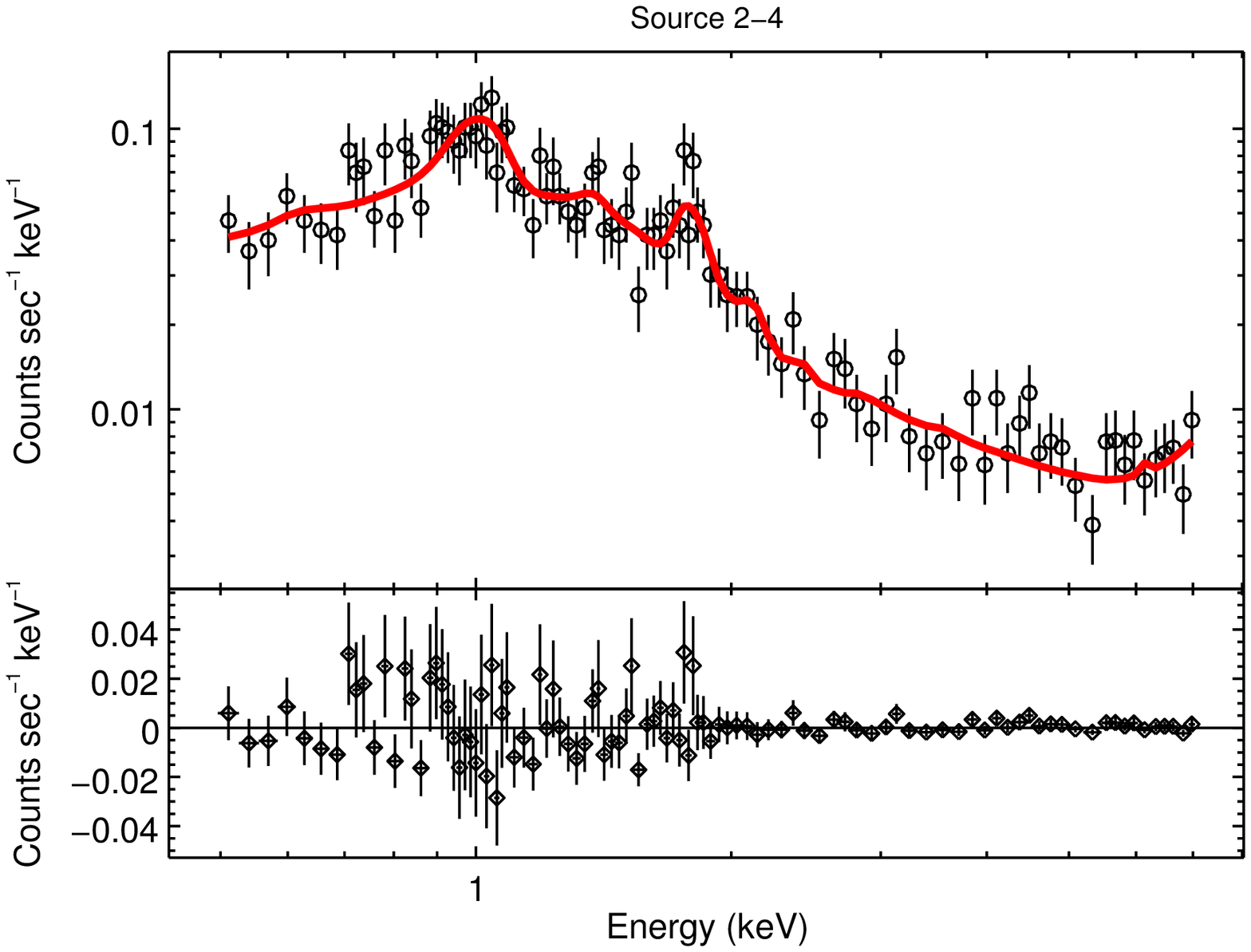}
\caption{Top panels show the observed X-ray spectra and spectral fits (with the abundance fixed at 0.5; red line) for the source 1 (left) and the source 2+3+4 (right)  regions (see \S\,\ref{sec:specfit}). Bottom panels show the differences between the observed data and the model.}
\label{spec_src}
\end{center}
\end{figure*}

\begin{figure*}[tbp]
\begin{center}
\includegraphics[width=8.9cm]{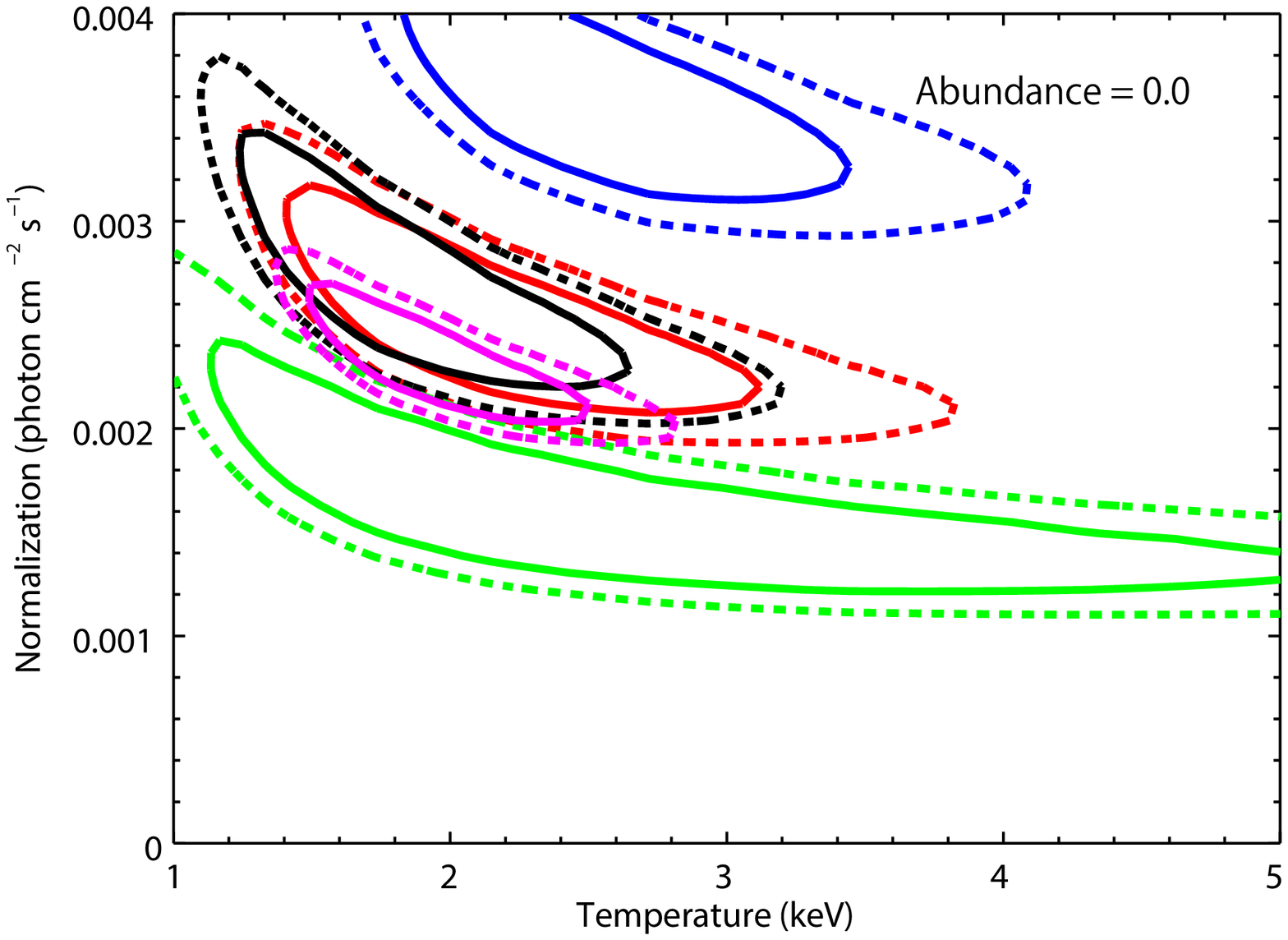}
\includegraphics[width=8.9cm]{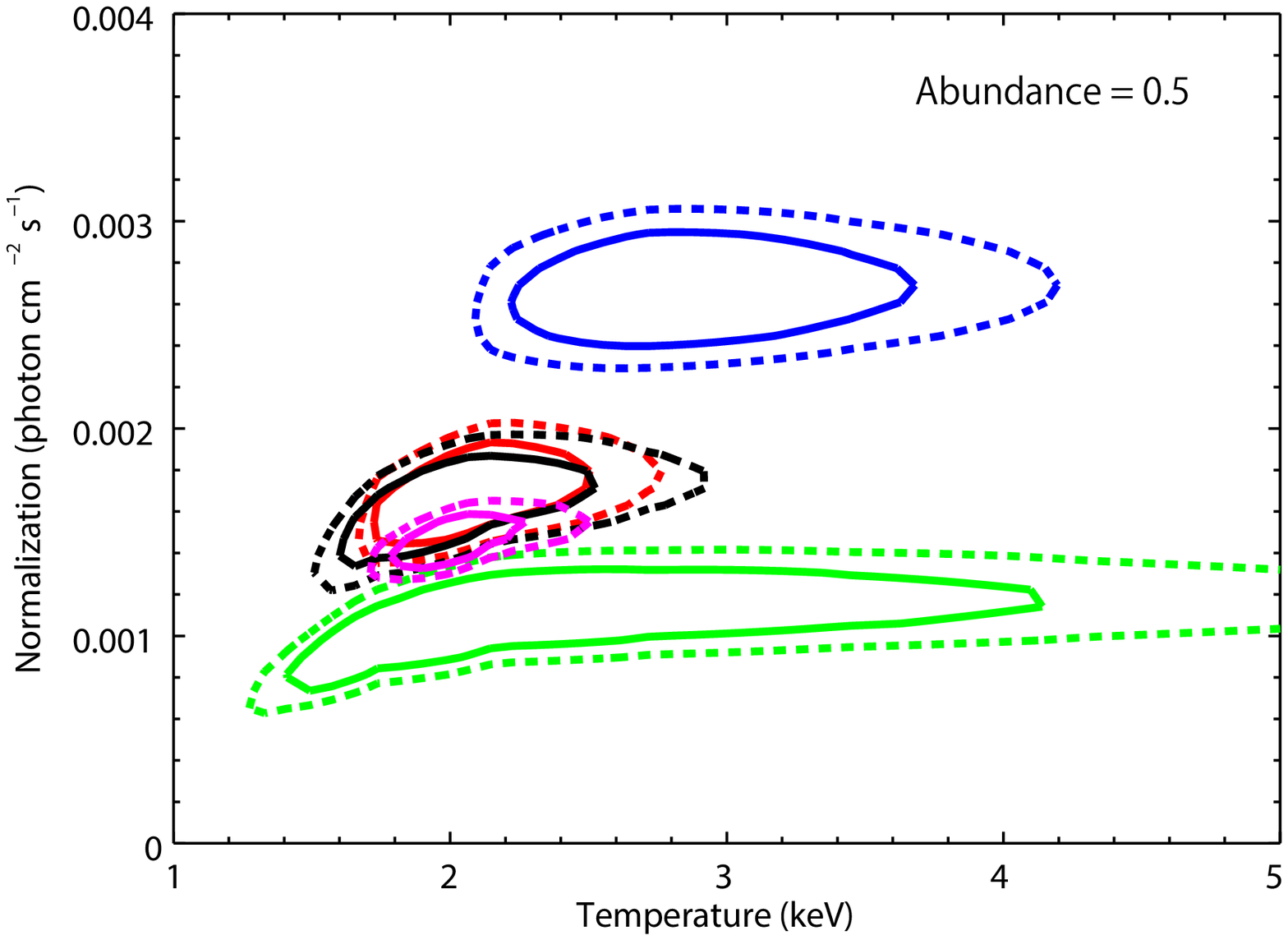}
\includegraphics[width=8.9cm]{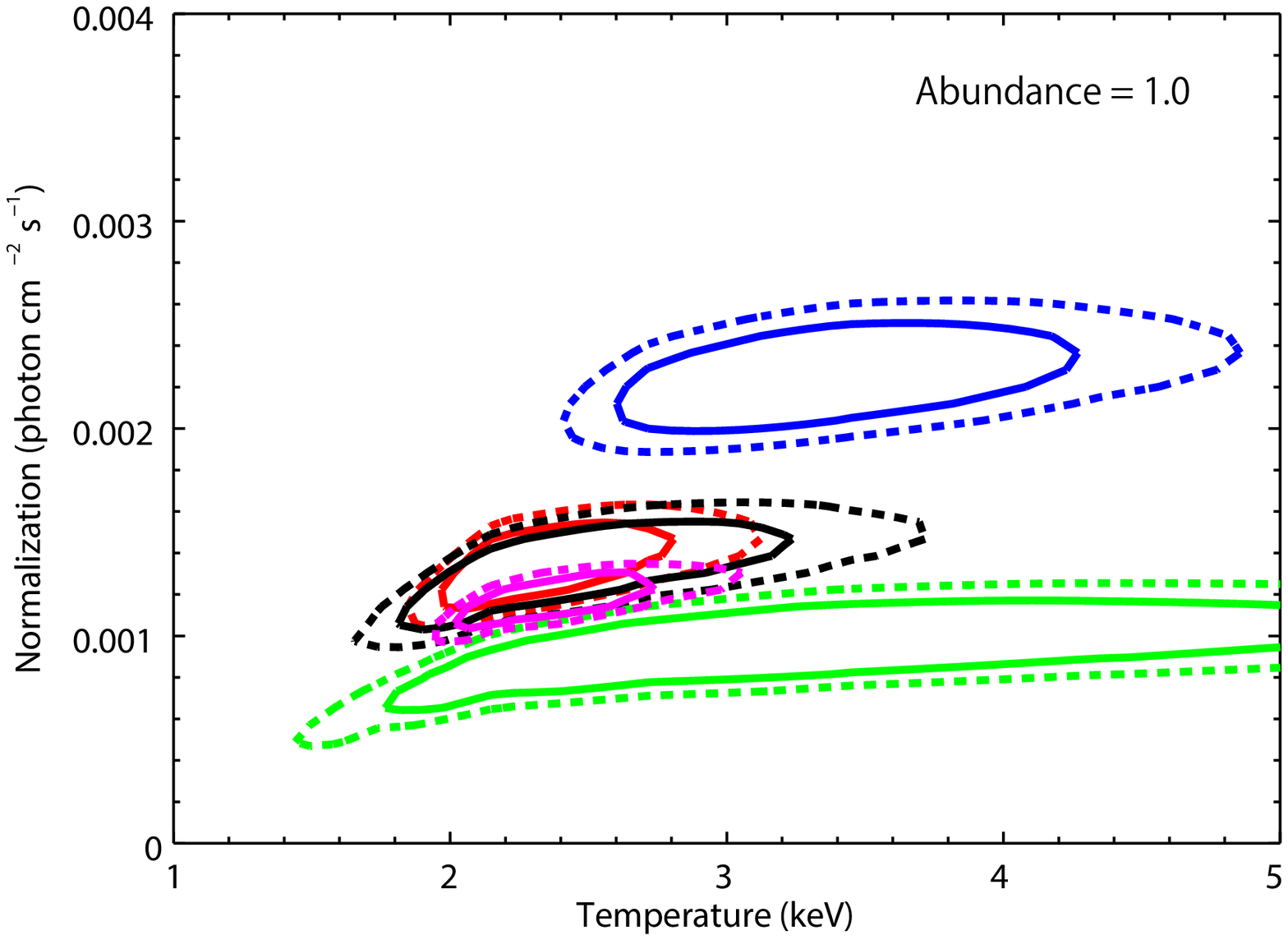}
\includegraphics[width=8.9cm]{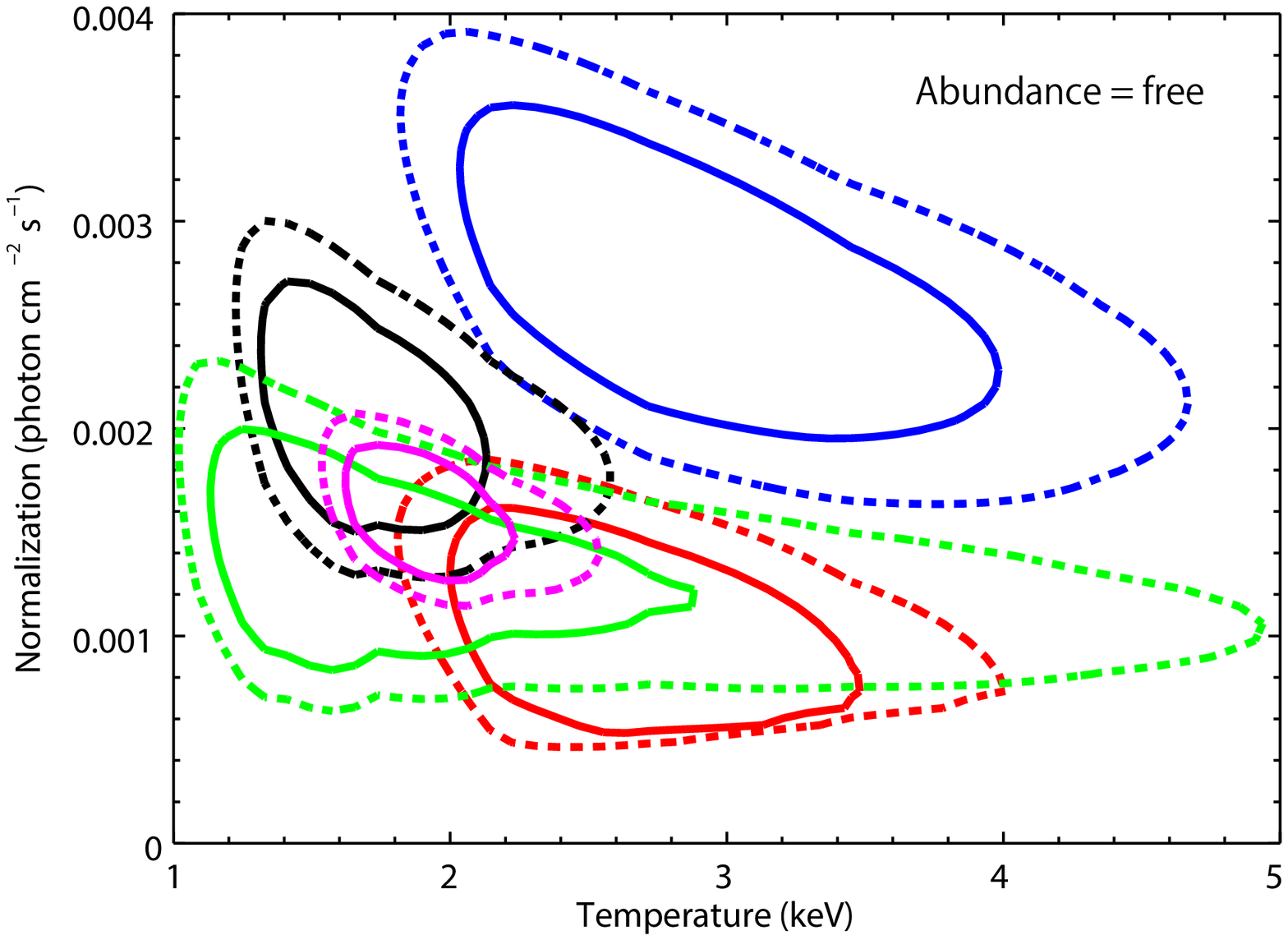}
\caption{Contour plots of temperature versus normalization corresponding to the spectral fits with different abundance values: 0.0 (top left), 0.5 (top right), 1.0 (bottom left) and free (bottom right). In each panel, the solid and dashed curves denote the 68$\%$ and 90$\%$ confidence levels, respectively, for the northern (black), eastern (red), southern (green), and western (blue) parts of the Abell\,578 cluster as discussed in \S\,\ref{sec:specfit} as shown in Figure\,\ref{reg_spec} with the same color-coding. The contours for the combined east+north+south region are indicated in magenta.}
\label{cont}
\end{center}
\end{figure*}

X-ray spectra and spectral fits for the source 1 and the combined source 2+3+4 regions are shown in Figure\,\ref{spec_src}; the model parameters for the abundance $A=0.5$ are listed in Table\,\ref{par_spec_src}. The total counts in the 0.5--7.0\,keV energy band were $838$ for source\,1, $884$ for source\,2, $880$ for source\,3, $773$ for source\,4, and $2537$ for source\,2+3+4. The corresponding contour plots of temperature versus normalization are shown in Figure \ref{cont}. For all the considered abundance values, source 1 (blue) is located at the upper right in the contour plots, which implies that the cluster gas around 4C\,+67.13 is indeed denser and hotter than that in the other selected regions, consistent with shock heating due to the expanding radio lobes. The reduced {\it C}-stat of the spectral fit for source 1 were 1.44, 1.33, 1.36 and 1.37 for the abundance 0, 0.5, 1 and free, respectively. Also, for the combined source 2+3+4, the reduced {\it C}-stat is best at the abundance of 0.5; hereafter we adopt the fitting results with the abundance fixed at this value. The temperature ratio of source 1 to the combined source 2+3+4 is then $1.37^{+0.30}_{-0.18}$, and the density ratio is $1.36^{+0.06}_{-0.06}$. These values correspond to shock Mach numbers of $\mathcal{M}_{sh} \simeq 1.38$ and 1.24, respectively.

\subsubsection{4C\,+67.13 Nucleus}
\label{S-nucleus}

Finally, we analyzed the X-ray spectrum of the active nucleus in 4C\,+67.13. The spectrum was extracted from a circular region with $1.5\arcsec$ radius while the background spectrum was extracted from the annular region from $3\arcsec$ to $11\arcsec$. The total number of counts in the 0.5--7.0\,keV energy band is $59$ for the source region and $88$ for the background region.

The background spectrum was fitted by the model used in the fitting of background A plus an APEC model moderated by the Galactic absorption. The normalizations of the instrumental background, as well as of the CXB and GH, were fixed at the values obtained in section \ref{sec:specfit}, and scaled by the areas of the extraction regions. Abundance was fixed at 1.0, and the redshift was fixed at 0.0866. The best-fit model parameters obtained are listed in\,Table \ref{par_core}. Note that the emerging temperature of 3.0\,keV is consistent with that obtained in spectral fitting of the source 1 region in \S\,\ref{sec:specfit}.

\begin{table*}[tbp]
\caption{Best-fit model X-ray parameters for the sources 1--4}
\begin{center}
\begin{tabular}{lccccc}
\hline\hline
Parameter & \multicolumn{5}{c}{Value}\\
\hline
& source 1 & source 2 & source 3 & source 4 & source 2+3+4\\
& West & East & North & South & North \& East \& South\\
\hline
Abundance & 0.5 (fix) & 0.5 (fix) & 0.5 (fix) & 0.5 (fix) & 0.5 (fix)\\
$kT$ [keV]  & $2.73^{+0.57}_{-0.31}$ & $2.04^{+0.26}_{-0.22}$ & $1.95^{+0.21}_{-0.26}$ & $1.98^{+1.23}_{-0.46}$ & $2.00^{+0.13}_{-0.14}$\\
norm\footnotemark[1] [$10^{-3}$] & $2.67^{ +0.19}_{-0.18}$&  $1.69^{ +0.17}_{-0.16}$ & $1.62^{+0.17}_{-0.16}$ & $1.07^{ +0.18}_{-0.24}$ & $1.46^{+0.09}_{-0.09}$\\
density [$10^{-3}~\mathrm{cm^{-3}}$] & $1.71^{+0.06}_{-0.06}$ & $1.36^{+0.07}_{-0.07}$ & $1.33^{+0.07}_{-0.07}$ & $1.08^{+0.09}_{-0.12}$ & $1.26^{+0.04}_{-0.04}$\\
reduced {\it C}-stat/dof & $1.33 / 37$ & $1.15 / 38$ & $0.79 / 39$ & $0.99 / 34$ & $1.19 / 100$\\
\hline
\end{tabular}
\footnotetext[1]{normalization of APEC model}
\end{center}
\label{par_spec_src}
\end{table*}

In the first attempt, the source (4C\,+67.13 nucleus) spectrum was fitted by a cosmologically redshifted power-law model with the Galactic absorption. The background model was fixed and the obtained best-fit parameters are listed in Table\,\ref{par_core}. In this case, the photon index of the power-law model obtained is quite large ($\Gamma=2.8$), and this model does not provide a very good fit ({\it C}-stat\,$=12.4/10$) due to a prominent excess around 1\,keV in the data. We have therefore added an additional thermal component to the fitted source spectrum \citep[consistently with the analysis results for LINER sources in general; see][]{you2011}, obtaining this time an acceptable fit with {\it C}-stat\,$=4.2/8$ (see Table\,\ref{par_core}) and a more reasonable photon index of $\Gamma=2.1$. The resulting temperature of the additional thermal component is $0.96$\,keV and the normalization of the power-law component is $1.8\times10^{-6}$\,photon\,keV$^{-1}$\,cm$^{-2}$\,s$^{-1}$. The total nuclear 2--10\,keV flux in the source rest frame is $F_{\mathrm{2-10\,keV}}\simeq3.4\times10^{-15}$\,erg\,cm$^{-2}$\,s$^{-1}$, corresponding to the isotropic luminosity of $L_{\mathrm{2-10\,keV}}\simeq6.2\times10^{40}$\,erg\,s$^{-1}$.
This is consistent with the bolometric accretion-related luminosity $L_{\rm nuc}$ estimated in \S\,\ref{S-opt} above, assuming the bolometric correction factor of the order of $L_{\rm nuc}/L_{\mathrm{2-10\,keV}} \sim 100$, as in fact expected \citep{vas07,ho08}.

\section{Discussion and Conclusions}

The Abell\,578 system provides an interesting insight into the widely debated cluster formation processes. Velocity measurements for the member galaxies presented by \citet{gom97} indicated that this system is not relaxed, and consists of the two interacting sub-clusters. Our analysis of newly obtained {\it Chandra} data allowed us to confirm their results based on ROSAT PSPC data for the dominant sub-cluster and to quantify them robustly. In particular, we found the ellipticity parameter for the cluster X-ray atmosphere $\epsilon \simeq 0.3$ (i.e., elongated east-west), and an offset of about $r_x \simeq 60$\,kpc between the BCG which hosts the 4C\,+67.13 radio galaxy and the centroid of the cluster X-ray emission (inferred from the two-dimensional fitting of the surface brightness distribution of the diffuse emission component). Similar displacements have been found in a substantial number of galaxy clusters \citep[e.g.,][]{pat06,haa10,has14}. Recently, \citet{lau14} demonstrated that the distribution of the projected spatial offsets is a steep power-law function of cluster radius with a median value of about 10\,kpc. The displacement we found for Abell\,578/4C\,+67.13 is therefore larger than in most clusters although not as extreme as in some other cases (such as, e.g., Coma cluster with $r_x \simeq 260$\,kpc), and is also smaller than the Abell\,578 cluster core radius $r_C \simeq 170$\,kpc (evaluated here for the assumed $\beta$-profile with best-fit index $\beta \simeq 0.7$). 

The position of the BCG with respect to the X-ray centroid of the cluster is considered a measure of the evolutionary state of a system. In particular, these offsets imply deviations from the dynamical equilibrium at earlier stages of a cluster formation via hierarchical merging \citep[see, e.g.,][]{mar14}. In addition, the presence of the offsets generally supports the idea that BCGs are formed via galaxy mergers in one of the infalling groups prior to the cluster assembly \citep[see the discussion in, e.g.,][]{mer85,dub98,kat03}. In this context, we note that the BCG studied here is a member of the galaxy pair CGPG 0719.8+6704 ($R$-band magnitudes of 13.6 and 14.8), with the projected separation of about 10\,kpc, and with no or only little young stellar populations. The performed optical spectroscopy allowed us also to estimate the black hole masses (from the observed velocity dispersion) for the two members of the pair as $M_{\rm BH} \sim 10^9 \, M_\odot$ and $3 \times 10^8 \, M_\odot$, respectively. The analysis of the emission lines indicated moreover that both galactic nuclei are active, albeit at a low level (LINER-type AGN), with the corresponding nuclear luminosities of $L_{\rm nuc} \sim 10^{43}$\,erg\,s$^{-1}$. However, only the brighter member of the pair is radio-loud. 

The accretion rate for the 4C\,+67.13 host is relatively low, namely $\dot{M}_{\rm acc}$~$= L_{\rm nuc}/ \eta_d \, c^2$~$\sim 0.01\,M_{\odot}$\,yr$^{-1}$~$\sim 5 \times 10^{-4}\,\dot{M}_{\rm Edd}$, assuming the radiative efficiency of the accretion disk $\eta_d \simeq 0.02$ \citep[following][]{sha07}. This is comparable to the accretion rate derived by \citet{sta14} for the analogous radio galaxy PKS\,B1358$-$113 located at the center of the Abell\,1836 cluster. Despite such a limited accretion rate, both AGN are able to launch luminous jets surrounded by FR\,II-type radio lobes. Interestingly, the \emph{maximum} jet kinetic power in the source studied here, for the given accretion rate as estimated above, reads as $L_{j/{\rm (max)}} \simeq 3 \, \dot{M}_{\rm acc} c^2 \sim 2 \times 10^{45}$\,erg\,s$^{-1}$ \citep[see][]{mck12}, which indeed is in the range of the jet kinetic luminosities typically derived for ``classical'' FR\,IIs \citep[e.g.,][]{mac07}.

We note, on the other hand, that it is quite plausible that the nuclear accretion rate in 4C\,+67.13 has dropped significantly since the onset of the jet activity in the source, i.e. during the last $\sim 1-10$\,Myr. And in fact, the lack of prominent hotspots at the edges of the radio lobes could be considered as an indication for a recently quenched/suppressed jet production in the system (due to a sudden decrease in the accretion rate), since jet termination shocks in FR\,II-type radio galaxies are expected to die out almost immediately (i.e., on the timescales of $\sim 10^4-10^5$\,yr) after the supply of the jet momentum ceases \citep[see in this context the discussion in][and references therein]{car96}. If correct, this could suggest in general a highly modulated jet duty cycle of BCGs, consisting of short but intense (``FR\,II-like'') episodes of the enhanced jet activity, separated by the extended (``FR\,I-like'') periods of a quiescence/low-level activity.

\begin{table}[tbp]
\caption{Best-fit model parameters for the 4C\,+67.13 nucleus}
\begin{center}
\begin{tabular}{lcc}
\hline\hline
Parameter & \multicolumn{2}{c}{Value}\\
\hline
\multicolumn{3}{c}{Source region}\\
\hline
Model                                   & power-law                     & power-law+apec \\
photon index                            & $2.8^{+0.3}_{-0.3}$   & $2.1_{-0.4}^{+0.5}$\\
norm\footnotemark[1] [$10^{-6}$] & $4.5^{+0.6}_{-0.6}$       &$1.8^{+1.1}_{-1.0}$\\
kT [keV]                                           & ---                                  &$0.96^{+0.15}_{-0.12}$\\
norm\footnotemark[2] [$10^{-6}$] & ---                            &$2.0^{+0.8}_{-0.8}$\\
reduced {\it C}-stat/dof                   & $1.24/10$                    &$0.52/8$\\
\hline
\multicolumn{3}{c}{Background region}\\
\hline
Abundance & \multicolumn{2}{c}{1.0 (fix)}\\
kT [keV] & \multicolumn{2}{c}{$3.01^{+2.02}_{-0.94}$}\\
norm\footnotemark[2] [$10^{-6}$] & \multicolumn{2}{c}{$51.3^{+9.6}_{-9.4}$}\\
reduced {\it C}-stat/dof & \multicolumn{2}{c}{$1.26/6$}\\
\hline
\end{tabular}
\footnotetext[1]{Normalization of the power-law model in units of $\mathrm{photons~keV^{-1}~cm^{-2}~s^{-1}}$ at 1~keV in the source rest frame.}
\footnotetext[2]{Normalization of the APEC model.}
\end{center}
\label{par_core}
\end{table}

In this context, we note that the existing scaling relations between the jet kinetic power and the lobes' radio luminosity for radio galaxies located at the centers of galaxy clusters have been derived assuming a relatively slow (sonic) expansion of the jet cocoons in a pressure equilibrium with the surrounding cluster gas \citep{bir08,cav10,osu11}. When applied to 4C\,+67.13, these relations give in particular $L_{j/{\rm (rad)}} \sim (2-7) \times 10^{44}$\,erg\,s$^{-1}$. The main assumption involved here may however be justified only in the case of FR\,I systems, but not FR\,II-type radio sources which likely expand with supersonic velocities; with such a rapid expansion, leading to the formation of bow-shocks in the surrounding intracluster medium, the jet kinetic power should be therefore larger than the corresponding value of $L_{j/{\rm (rad)}}$.

Our {\it Chandra} data do not provide any \emph{direct} evidence for the presence of a shock driven in the center of Abell\,578 by a supersonic expansion of the 4C\,+67.13 lobes, but this may only be due to the limited photon statistics. Instead, our detailed analysis of the available X-ray maps did reveal that the cluster gas in the vicinity of 4C\,+67.13 seems to be compressed by a factor of about 1.4  (i.e., the gas number density increased by this factor) and heated (from $kT \simeq 2.0$\,keV up to $2.7$\,keV), consistent with a weak, $\mathcal{M}_{sh} \sim 1.3$ shock driven by the expanding jet cocoon. Analogous weak shocks have been found in much deeper {\it Chandra} and XMM-{\it Newton} exposures of several rich clusters \citep[e.g.,][]{nul05,fab06,wil06,git07,wis07,sim07,rey08,lal10,bla11,cav11,cro11}, and further, often tentative detections have been reported for a number of other (poorer) systems, including PKS\,B1358$-$113 \citep[][and references therein]{sta14}.

One should keep in mind, at the same time, that the aforementioned limited photon statistics following from a rather short {\it Chandra} exposure of a relatively low-luminosity system, 
precludes us from making definite statements on the precise gas density and temperature structure in the central parts of Abell\,578. In particular, it is plausible that the elongated X-ray core of the cluster, together with the large offset between the position of the BCG and the X-ray centroid inferred from the modeling of the diffuse emission component, are related instead to the sloshing of the intracluster medium triggered by the ongoing merger processes \citep[e.g.,][]{tit05,zuh10}. Only very deep follow-up observations of the studied system can resolve this issue by revealing directly the surface brightness and temperature discontinuities consistent with ``cold fronts'' --- expected in the case of a gas sloshing --- rather than weak shocks \citep[see in this context][]{mar07}.

Altogether, the multifrequency data gathered here for the Abell\,578 cluster reveal an insightful snap-shot of a complex system in formation. At its present stage, it appears that the cluster galaxies and intracluster gas are falling onto the center of the cluster potential, evolving slowly toward a dynamical equilibrium, while the cluster atmosphere is heated gently by the radio jets/lobes produced with high efficiency in the low-accretion rate nucleus of the brightest cluster galaxy. The overall cluster luminosity $L_{\rm 0.5-7.0\,keV} \simeq 2 \times 10^{43}$\,erg\,s$^{-1}$ and temperature $kT \gtrsim 2$\,keV are on the other hand unexceptional, being consistent with the X-ray luminosity-temperature correlation established for clusters and groups of galaxies \citep[see, e.g.,][]{sun12}.

\section*{Acknowledgments}
K.~H. is supported by the Japan Society for the Promotion of Science (JSPS) Research Fellowship for Young Scientists.
\L .~S. was supported by Polish NSC grant DEC-2012/04/A/ST9/00083.
Support for A.~S. was provided by NASA contract NAS8-03060.
Work by C.~C.~C. at NRL is supported in part by NASA DPR S-15633-Y. 
A.~Sz. and G.~M. were supported by Chandra grant GO0-11144X.
The authors thank the anonymous referee for her/his constructive comments which helped to improve the paper.

{}

\end{document}